ON THE RECTILINEAR MOTION OF THREE BODIES MUTUALLY ATTRACTING EACH OTHER
(The three-body problem on a straight line)
*E327 -- De motu rectilineo trium corporum se mutuo attrahentium*

Translated and Annotated
by Sylvio R. Bistafa*
**********
March 2019

Foreword

Euler wrote a number of papers concerned with the three-body problem: *Considerationes de motu corporum coelestium* (Considerations on the motion of celestial bodies, E304, 1766); *De motu rectilineo trium corporum se mutuo attrahentium* (On the rectilinear motion of three bodies mutually attracting each other, E327, 1767); and *Considérations sur le problèm des trois corps* (Considerations on the problem of three bodies, E400, 1770). This is an annotated translation from Latin of E327. In this publication, Euler considers three bodies lying on a straight line, which are attracted to each other by central forces inversely proportional to the square of their separation distance (inverse-square law). Although not explicitly mentioned by Euler, this is an exact solution of three bodies that move around the common center of mass and always line up. The solution given by Euler could represent a hypothetical situation of Sun, Earth and Moon in perpetual alignment in syzygy, for which the parameter that controls the distances among the planets was found to be given by a quintic function.

On the rectilinear motion of three bodies mutually attracted to each other

Author
L. Euler

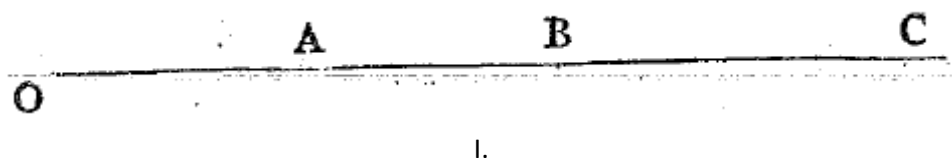

I.

Let $A, B, C$ be the masses of three bodies such that their distances to a fixed point $O$ at a given instant of time $t$ is given by

$$OA = x, OB = y \text{ and } OC = z$$

where, in fact, it is assumed that $y > x$ and $z > y$. Hence, the principles of motion give these three equations:

---
* Corresponding address: sbistafa@usp.br

$$\text{I.} \quad \frac{ddx}{dt^2} = \frac{B}{(y-x)^2} + \frac{C}{(z-x)^2}; \quad [1]$$

$$\text{II.} \quad \frac{ddy}{dt^2} = \frac{-A}{(y-x)^2} + \frac{C}{(z-y)^2} \quad [2]$$

$$\text{III.} \quad \frac{ddz}{dt^2} = \frac{-A}{(z-x)^2} - \frac{B}{(z-y)^2} \quad [3]$$

whence two integrable equations are easily derived: the first [integral]

$$Adx + Bdy + Cdz = Edt \quad [4]$$

which upon integration results in

$$Ax + By + Cz = Et + F;^1 \quad [5]$$

and the second [integral]

$$\frac{Adx^2 + Bdy^2 + Cdz^2}{dt^2} = G + \frac{2AB}{y-x} + \frac{2AC}{z-x} + \frac{2BC}{z-y}.^2 \quad [6]$$

Whence, because we lack a third integral equation, very little is possible to conclude about the movement.

2. Let us set $x = y - p$ and $z = y + q$, such that $p$ and $q$ are positive quantities; and the first integral [Eq.5] gives:

$$(A + B + C)y - Ap + Cq = Et + F \quad [7]$$

and thus

$$y = \frac{Ap - Cq + Et + F}{A + B + C}; \quad dy = \frac{Adp - Cdq + Edt}{A + B + C} \quad [8\,a, b]$$

$$x = \frac{-(B+C)p - Cq + Et + F}{A + B + C}; \quad dx = \frac{-(B+C)dp - Cdq + Edt}{A + B + C} \quad [9\,a, b]$$

$$z = \frac{Ap + (A+B)q + Et + F}{A + B + C}; \quad dx = \frac{Adp + (A+B)dq + Et}{A + B + C}. \quad [10\,a, b]$$

Whence the second integral [Eq. 6] assumes the following form:

$$\frac{A(B+C)dp^2 + C(A+B)dq^2 + 2ACdpdq + EEdt^2}{(A+B+C)dt^2} = G + \frac{2AB}{p} + \frac{2AC}{p+q} + \frac{2BC}{q}, \quad [11]$$

whence arises one integral equation[3]

$$\frac{B(Adp^2 + Cdq^2) + AC(dp + dq)^2}{(A+B+C)dt^2} = G + \frac{2AB}{p} + \frac{2AC}{p+q} + \frac{2BC}{q}, \quad [12]$$

---

[1] Upon the multiplication of $I.$[Eq. 1] by $A$; $II.$ [Eq. 2] by $B$; $III.$ [Eq. 3] by $C$, and taking the sum over the three equations yields $Adx + Bdy + Cdz = Edt$, which upon integration gives $Ax + By + Cz = Et + F$, where $E$ and $F$ are constants of integration.

[2] From $I. \frac{Adx}{dt} = \int \frac{AB}{(y-x)^2} dt + \int \frac{AC}{(z-x)^2} dt$, which upon multiplication by $\frac{dx}{dt}$ gives $\frac{Adx^2}{dt^2} = \int \frac{AB}{(y-x)^2} dx + \int \frac{AC}{(z-x)^2} dx$, and then $\frac{Adx^2}{dt^2} = \frac{AB}{y-x} + \frac{AC}{z-x}$. And similarly for II. and III., yielding: $\frac{Bdy^2}{dt^2} = \frac{AB}{y-x} + \frac{BC}{z-y}$, and $\frac{Cdz^2}{dt^2} = \frac{AC}{z-x} + \frac{BC}{z-y}$. Taking the sum over the three equations and introducing the constant of integration $G$ yields $\frac{Adx^2 + Bdy^2 + Cdz^2}{dt^2} = G + \frac{2AB}{y-x} + \frac{2AC}{z-x} + \frac{2BC}{z-y}$.

[3] In the original manuscript, these three last lines have been misplaced at the end of § 3.

and observing that the last term $EE$ is included into $G$.

3. Let us make the same proper substitutions into the first second order differential [difference - of - differentials] equations [Eqs. 1 and 3], which now result in two [equations]:

$$\frac{-(B+C)ddp - Cddq}{(A+B+C)dt^2} = \frac{B}{pp} + \frac{C}{(p+q)^2} \qquad [13]$$

$$\frac{Addp + (A+B)ddq}{(A+B+C)dt^2} = \frac{-A}{(p+q)^2} - \frac{B}{qq}, \qquad [14]$$

which [by subtracting Eq. 13 from Eq. 12] results in

$$\frac{ddp + ddq}{dt^2} = \frac{-A-C}{(p+q)^2} - \frac{B}{pp} - \frac{B}{qq}. \qquad [15]$$

And then, each element $ddp$ and $ddq$ can be expressed separately in the following way

$$1^{\underline{o}}. \quad \frac{ddp}{dt^2} = \frac{-A-B}{pp} - \frac{C}{(p+q)^2} + \frac{C}{qq} \qquad [16]$$

$$2^{\underline{o}}. \quad \frac{ddq}{dt^2} = \frac{A}{pp} - \frac{A}{(p+q)^2} - \frac{B \mp C}{qq}. \qquad [17]$$

4. Since the solution has been reduced to two differential equations involving $p$, $q$ and $t$ we should expect that significant advantage is to be obtained, if it were possible to reduce these equation to two others of first order only. This is a unique technique that I have discovered which can be applied in the following manner. I put $q = pu$, and the two differential equations [Eqs. 15 and 16] are represented as:

$$d\left(\frac{dp}{dt}\right) = \frac{dt}{pp}\left(-A - B - \frac{C}{(u+1)^2} + \frac{C}{uu}\right) \qquad [18]$$

$$d\left(\frac{udp + pdu}{dt}\right) = \frac{dt}{pp}\left(A - \frac{A}{(u+1)^2} - \frac{B \mp C}{uu}\right). \qquad [19]$$

Now the trick consists in the following substitution $\frac{dp}{dt} = \frac{r}{\sqrt{p}}$ and $\frac{dq}{dt} = \frac{udp+pdu}{dt} = \frac{s}{\sqrt{p}}$; because it will expose that for these substitutions, the two variables $p$ and $t$ can be eliminated from the calculations, such that only these three [variables] $r$, $s$ and $u$ are to be determined by their first differentials. Then, in particular, the equation that the integral was found above [Eq. 17] assumes a finite form which reads

$$\frac{B(Arr + Css) + AC(r+s)^2}{A+B+C} = Gp + 2AB + \frac{2AC}{u+1} + \frac{2BC}{u}, \qquad [20]$$

whose usefulness it will be possible to assess.

5. Since $\frac{dp}{dt} = \frac{r}{\sqrt{p}}$, then $dt = \frac{dp\sqrt{p}}{r}$, whence our second order differential [difference - of - differentials] equations [Eqs. 18 and 19] give

$$\frac{dr}{\sqrt{p}} - \frac{rdp}{2p\sqrt{p}} = \frac{dp}{pr\sqrt{p}}\left(-A - B - \frac{C}{(u+1)^2} + \frac{C}{uu}\right) \qquad [21]$$

$$\frac{ds}{\sqrt{p}} - \frac{sdp}{2p\sqrt{p}} = \frac{dp}{pr\sqrt{p}}\left(A - \frac{A}{(u+1)^2} - \frac{B \mp C}{uu}\right). \qquad [22]$$

or:

$$dr = \frac{rdp}{2p} + \frac{dp}{pr}\left(-A - B - \frac{C}{(u+1)^2} + \frac{C}{uu}\right) \qquad [23]$$

$$ds = \frac{sdp}{2p} + \frac{dp}{pr}\left(A - \frac{A}{(u+1)^2} - \frac{B \mp C}{uu}\right). \qquad [24]$$

Moreover, in particular, it will be considered that

$$udp + pdu = \frac{sdt}{\sqrt{p}} = \frac{sdp}{r}, \qquad [25]$$

such that $\frac{dp}{p} = \frac{rdu}{s-ru}$ which when substituted [into Eqs. 23 and 24] gives

$$dr(s - ru) = \frac{1}{2}rrdu + du\left(-A - B - \frac{C}{(u+1)^2} + \frac{C}{uu}\right) \qquad [26]$$

$$ds(s - ru) = \frac{1}{2}rsdu + du\left(A - \frac{A}{(u+1)^2} - \frac{B \mp C}{uu}\right), \qquad [27]$$

which when combined give:

$$\frac{1}{2}r(rds - sdr) + ds\left(-A - B - \frac{C}{(u+1)^2} + \frac{C}{uu}\right) - dr\left(A - \frac{A}{(u+1)^2} - \frac{B \mp C}{uu}\right) = 0. \qquad [28]$$

6. We see that we have two first-order differential equations [Eqs. 26 and 27] among three variables $r$, $s$ and $u$, whence if it were possible to determine $r$ and $s$ in terms of of $u$, and then one would have the complete solution of the problem. Thence, in fact, $p$ would become known from the formula $\frac{dp}{p} = \frac{rdu}{s-ru}$, and hence furthermore $q = pu$. Thereafter, the particular time $t$ would be given from the equation $dt = \frac{dp\sqrt{p}}{r} = \frac{pdu}{s-ru}$; and finally, for a given time $t$, the distances $x, y, z$ would be obtained as given in § 2.

7. Since the two differential equations [Eqs. 26 and 27] found are

$$dr(s - ru) = \frac{1}{2}rrdu + du\left(-A - B - \frac{C}{(u+1)^2} + \frac{C}{uu}\right)$$

$$ds(s - ru) = \frac{1}{2}rsdu + du\left(A - \frac{A}{(u+1)^2} - \frac{B - C}{uu}\right),$$

then, it is clear that both are satisfied by taking the quantity $u$ constant and $s - ru = 0$, whence a particular solution is obtained. If $u = \propto$ and $s = \propto r$ [then from Eq. 28 we have that:]

$$-(A + B)\propto - \frac{C\propto}{(\propto +1)^2} + \frac{C}{\propto} = A - \frac{A}{(\propto +1)^2} - \frac{B \mp C}{\propto\propto}, \qquad [29]$$

or

$$0 = A\left(\propto +1 - \frac{1}{(\propto +1)^2}\right) + B\left(\propto - \frac{1}{\propto\propto}\right) + C\left(\frac{\propto}{(\propto +1)^2} - \frac{1}{\propto} - \frac{1}{\propto\propto}\right), \qquad [30]$$

or else

$$0 = A\frac{[(\propto +1)^3 - 1]}{(\propto +1)^2} + \frac{B(\propto^3 - 1)}{\propto\propto} + \frac{C[\propto^3 - (\propto +1)^3]}{\propto\propto (\propto +1)^2}; \qquad [31]$$

hence,

$$C(1 + 3\alpha + 3\alpha\alpha) = A\alpha^3(\alpha\alpha + 3\alpha + 3) + B(\alpha+1)^2(\alpha^3 - 1). \quad [32]$$

Thus it is possible to determine the quantity $\alpha$ from this equation of the fifth degree :

$$(A+B)\alpha^5 + (3A+2B)\alpha^4 + (3A+B)\alpha^3 - (B+3C)\alpha^2 - (2B+3C)\alpha - B - C = 0.^4 \quad [33]$$

Thence, truly from the relation between $r$ and $p$ [Eq. 21] this equation is obtained

$$dr = \frac{rdp}{2p} + \frac{dp}{pr}\left(-A - B - \frac{C}{(\alpha+1)^2} + \frac{C}{\alpha\alpha}\right) \quad [34]$$

or by putting $A + B + \frac{C}{(\alpha+1)^2} - \frac{C}{\alpha\alpha} = \frac{1}{2}D$, [then, from Eq. 34 we have]

$$2dr = \frac{dp}{p}\left(r - \frac{D}{r}\right) \quad \text{or} \quad \frac{dp}{p} = \frac{2rdr}{rr - D}, \quad [35\ a, b]$$

which [upon integration of Eq. 35b] gives

$$p = \beta(rr - D), \quad [36]$$

[where $\zeta$ is a constant of integration], and then [since $q = pu = p\alpha$]

$$q = \alpha\beta(rr - D), \quad [37]$$

and [since] $dt = \frac{dp\sqrt{p}}{r}$ or $dt = 2\beta dr\sqrt{\beta(rr - D)}$, hence

$$t = \beta r\sqrt{\beta(rr - D)} - \beta^2 D \int \frac{dr}{\sqrt{\beta(rr - D)}}. \quad [38]$$

8. This particular case, in which the solution succeeds, deserves to be unfolded carefully. Firstly, I observe that the value of $\alpha$, since it is obtained from an equation of the fifth degree, is unique and always a positive real quantity, because there is only one sign variation[5], and then, of course, there is no reason for any ambiguity; however, the value of this $\alpha$ can be seen to depend on the masses of the three bodies $A, B, C$. The number $\alpha$ having been found, we get the quantity $D = 2(A + B) - \frac{2C(2\alpha+1)}{\alpha\alpha(\alpha+1)^2}$, where it should be observed that the quantity $D$ can never vanish. In fact, if $D = 0$, then $B = \frac{C(2\alpha+1)}{\alpha\alpha(\alpha+1)^2} - A$, which, when substituted [into Eq. 32] would give:

$$C(1 + 3\alpha + 3\alpha\alpha) = A\alpha^3(\alpha\alpha + 3\alpha + 3) + \frac{C(2\alpha+1)(\alpha^3 - 1)}{\alpha\alpha} - A(\alpha+1)^2(\alpha^3 - 1), \quad [39]$$

or

$$\frac{C}{\alpha\alpha}(\alpha^4 + 2\alpha^3 + \alpha\alpha + 2\alpha + 1) = A(\alpha^4 + 2\alpha^3 + \alpha\alpha + 2\alpha + 1), \quad [40]$$

---

[4] Equation [28], from which the fifth-degree equation [33] for $\alpha$ is ultimately derived, is obtained from [26] and [27] by eliminating $du/(s - ru)$ between them. But, in the special solution that Euler considers, namely, $u = \alpha$, $s = \alpha r$, both sides of [26] and [27] become identically 0. In this case, therefore, [26] and [27] could be true, regardless of whether [28] holds or not. So, despite the ingenuity with which Euler investigates the consequences of the fifth-degree equation [33], it is not clear that his analysis is well-founded.

[5] Descartes' Rule of Signs states that if the terms of a single-variable polynomial with real coefficients are ordered by descending variable exponent, then the number of positive roots of the polynomial is either equal to the number of sign differences between consecutive nonzero coefficients, or is less than it by an even number.

and, therefore, $C = A \propto\propto$ and $B = \frac{A(2\propto+1)}{(\propto+1)^2} - A = \frac{-A\propto\propto}{(\propto+1)^2}$, and then, $B$ would be a negative mass, which is absurd. Even less possible is that the quantity $D$ could be negative. In fact, assuming [that $\frac{D}{2} = -\Delta$, where $\Delta$ is a positive quantity]:

$$B = \frac{C(2\propto+1)}{\propto\propto(\propto+1)^2} - A - \Delta, \quad [41]$$

it would then give [when substituted into Eq. 32]:

$$\frac{C}{\propto\propto} = A - \frac{\Delta(\propto+1)^2(\propto^3-1)}{\propto^4 + 2\propto^3 + \propto\propto + 2\propto + 1}, \quad [42]$$

hence [by isolating $A$ in the first term of Eq. 42, and substituting the resulting expression into Eq. 41]

$$B = \frac{C(2\propto+1)}{\propto\propto(\propto+1)^2} - \frac{C}{\propto\propto} - \frac{\Delta(\propto^5 + 3\propto^4 + 3\propto^3)}{\propto^4 + 2\propto^3 + \propto\propto + 2\propto + 1}, \quad [43]$$

and then, $B$ would be a much more negative quantity, since it is necessary that the value of $\propto$ itself be positive.

9. Then, since it is necessary that the quantity $D$ be positive, it can be assumed that $D = aa$, and if also the number $\propto$ is considered as given, therefore the masses of the three bodies will be obtained as [from Eq. 43, with $\Delta = \frac{aa}{2}$]:

$$B = \frac{(\propto^5 + 3\propto^4 + 3\propto^3)aa}{2(\propto^4 + 2\propto^3 + \propto\propto + 2\propto + 1)} - \frac{C}{(\propto+1)^2}; \quad [44]$$

and [from Eq. 42, with $\Delta = -\frac{aa}{2}$]:

$$A = \frac{C}{\propto\propto} - \frac{(\propto+1)^2(\propto^3-1)aa}{2(\propto^4 + 2\propto^3 + \propto\propto + 2\propto + 1)}, \quad [45]$$

from which it is necessary that the quantity $\frac{2C(\propto^4 + 2\propto^3 + \propto\propto + 2\propto + 1)}{\propto\propto(\propto+1)^2 aa}$ be bound between the limits $(\propto+1)^3 - 1$ and $\propto^3 - 1$. Then, once the quantity $aa$ with the number $\propto$ are introduced into the calculations, two cases should be examined, according to the sign of the quantity $\beta$ (positive or negative), which we shall examine separately.

Case I.

10. Be first $\beta = nn$, and then $p = nn(rr - aa)$ [from Eq. 36] and $q = \propto nn(rr - aa)$ [from Eq. 37], and by putting the constants $E$ and $F$ equal to zero, the locations of the three bodies $A, B, C$, whose center of gravity is now located in $O$, are defined by $r$ such that:

$$x = OA = \frac{-nn(rr-aa)}{A+B+C}(B + C + C\propto), \quad [46]$$

$$y = OB = \frac{nn(rr-aa)}{A+B+C}(A - C\propto), \quad [47]$$

$$z = OC = \frac{nn(rr-aa)}{A+B+C}(A + (A+B)\propto). \quad [48]$$

Yet, the relation between $r$ and the time $t$ is [Eq. 38]

$$t = n^3 r\sqrt{rr - aa} - n^3 aa \int \frac{dr}{\sqrt{rr - aa}}, \qquad [49]$$

or

$$t = n^3 r\sqrt{rr - aa} - n^3 aa \ln\left|\frac{r + \sqrt{rr - aa}}{\Delta}\right|. \qquad [50]$$

Assuming that the constant $\Delta = a$, then, for the time $t = 0$, $r = a$, meaning that all bodies are concentrated in the center of gravity $[O]$, whence they will be driven out with an almost infinite velocities, and then, these [distances] are similar to each other as the quantities of: $-B - C - C \propto, A - C \propto, A + (A + B) \propto$; also, with the passage of time $t$ the quantity $r$ increases even more; however, for any other time, the velocity of each of the bodies becomes known from the formula $\frac{dt}{dr} = 2n^3\sqrt{rr - aa}$. However, noting that the inter body distances preserve the same proportion.

Case II.

11. Be now $\beta = -nn$, and then $p = nn(aa - rr)$ [from Eq. 36] and $q = \propto nn(aa - rr)$ [from Eq. 37], and the locations of the bodies are defined by $r$ from:

$$x = OA = \frac{-nn(aa - rr)}{A + B + C}(B + (1+\propto)C), \qquad [51]$$

$$y = OB = \frac{nn(aa - rr)}{A + B + C}(A - C \propto), \qquad [52]$$

$$z = OC = \frac{nn(aa - rr)}{A + B + C}(A(\propto +1) + B \propto). \qquad [53]$$

On the other hand, for the time $t$ to be obtained, $t = 2n^3 dr\sqrt{(aa - rrr)}$,

or

$$t = n^3 r\sqrt{aa - rr} + n^3 aa \int \frac{dr}{\sqrt{aa - rr}}, \qquad [54]$$

hence,

$$t = n^3 r\sqrt{aa - rr} + n^3 aa \sin^{-1}\left(\frac{r}{a}\right). \qquad [55]$$

But, by putting $\sin^{-1}\left(\frac{r}{a}\right) = \phi$, such that $r = a\sin\phi$, then $t = n^3 aa(\phi + \sin\phi\cos\phi)$, and at any time, the inter distances are proportional to $\cos^2\phi$.[6] Whence, if at the beginning when $t = 0$, also $\phi = 0$, thus $r = 0$, and $\frac{dt}{dr} = 2n^3 a$, then the distances will be:

$$x = OA = \frac{-nnaa}{A + B + C}(B + (1+\propto)C), \qquad [56]$$

$$y = OB = \frac{nnaa}{A + B + C}(A - C \propto), \qquad [57]$$

$$z = OC = \frac{nnaa}{A + B + C}(A(\propto +1) + B \propto), \qquad [58]$$

---

[6] According to Eqs. 51-53, the distances are proportional to $(a^2 - r^2) = \left(a^2 - a^2\frac{r^2}{a^2}\right) = a^2(1 - \sin^2\phi) = a^2\cos^2\phi$.

and in that place the bodies are at rest. On the other hand, once it has been assumed that $\phi = 90º$, or after the time $t = n^3 aa \cdot 90º$ has elapsed, the bodies approach the center of gravity with infinity velocity.[7]

______________________

An Application by the Translator

For $A = B = C = 1$ unit, the fifth order polynomial [Eq.33] reduces to

$$2\alpha^5 + 5\alpha^4 + 4\alpha^3 - 4\alpha^2 - 5\alpha - 2 = 0,$$

which, by inspection, gives $\alpha = 1$ as the single positive root.

For Case I, and from Eqs. 46-48, the positions of these masses are given by $x = -nn(rr - aa)$, $y = 0$, and $z = nn(rr - aa)$, which show that one of the masses occupies the center of gravity, whereas the two other masses occupy symmetrical positions in relation to the center of gravity, and that these positions depend on $(rr - aa)$.

For Case II, and when $t = 0$, the motion begins in a symmetrical configuration of the masses in relation to the center of gravity, however, according to Eqs. 56-58, the positions of these masses are now given by $x = -nnaa$, $y = 0$, and $z = nnaa$. For any other instant of time, the positions of the masses will be given by $x = -nn(aa - rr)$, $y = 0$, and $z = nn(aa - rr)$, which show that the symmetrical configuration is preserved, however, the positions of the masses now depend on $(aa - rr)$ instead.

The conclusion is that for these particular cases, the configuration of the masses during the motion is such that one of the masses occupies the center of gravity, with the two other masses remaining on the same straight line, and moving symmetrically around the center of gravity of the system. This could represent a hypothetical situation of Sun, Earth and Moon in perpetual alignment in syzygy, for which the parameter that controls the distances among the planets was found to be given by a quintic function.

---

[7] Since $dt = -2nndr\sqrt{nn(aa - rr)}$, or $\frac{dr}{dt} = -\frac{1}{2n^3\sqrt{(aa-rr)}}$, which for $\phi = 90º$ $(r = a)$ gives $\frac{dr}{dt} = -\infty$.